\documentclass{article}
\usepackage{graphicx}
\usepackage{setspace}
\usepackage{bm}

\usepackage{graphicx}
\usepackage{citesort}
\usepackage{epic,eepic}
\usepackage{graphics}
\usepackage{epsfig}
\usepackage{pifont}
\usepackage{subfigure}
\usepackage{amsmath}

\newcommand{\ee}{\end{equation}}
\newcommand{\be}{\begin{equation}}
\newcommand{\eea}{\end{eqnarray}}
\newcommand{\bea}{\begin{eqnarray}}

\newcommand{\pat}{\partial}

\begin{document}
\title{A note on the effective slip properties for microchannel flows with 
ultra-hydrophobic surfaces}
\author{
{M. Sbragaglia $^{1}$  and A. Prosperetti $^{1,2}$}\\ 
{$^1$ Faculty of Applied Sciences, IMPACT, and Burgerscentrum,}\\ 
{University of Twente, AE 7500 Enschede, The Netherlands}\\
{$^2$ Department of Mechanical Engineering}\\ 
{The Johns Hopkins University, Baltimore MD 21218, USA}}


\maketitle
\begin{abstract}{A type of super-hydrophobic surface consists of a solid 
plane boundary with an array of grooves which, due to the effect of surface tension,  prevent a complete wetting of the wall. The effect is greatest when the grooves are aligned with the flow. The pressure difference between the liquid and the gas in the grooves causes a curvature of the liquid surface resisted by 
surface tension. The effects of this surface deformation are studied in this 
paper. The corrections to the effective slip length produced by the
curvature are analyzed theoretically and a comparison with available data and
related mathematical models is presented.}
\end{abstract}
\maketitle


\section{Introduction}

Large pressure drops are necessary to cause liquid flow in micro- and 
nano-channels. The small values of naturally occurring slip lengths on 
hydrophobically coated surfaces, typically of the order of some nanometers 
\cite{laugarev06,cheng02,vino05,zhu02}, are in most cases 
insufficient to significantly affect the required pressure gradients, as 
also confirmed by some theoretical analysis 
\cite{barrat99,Cottin04,PRLsbragag06}. 
With the recent increasing interest in such systems  
\cite{hotai98,tabebook03}, efforts to reduce these large pressures have 
been made. One promising avenue are the so-called super-hydrophobic surfaces 
\cite{ou04,ou05,tabe06,bico99,oner00,wata99}, in which  the wall stress 
is decreased by reducing the liquid-solid contact area 
\cite{laugarev06,tabebook03}. The idea is to cover the surface of interest 
with structures such as posts, grooves, or others, over which the liquid 
surface remains suspended due to the effect of surface tension. 
The effect is equivalent to the introduction of an effective slip length of
a magnitude comparable with the size of the geometrical features on the 
surface. This 
arrangement has been studied by several authors both experimentally 
\cite{ou04,ou05,tabe06,choi06} and theoretically \cite{barrat99,PRLsbragag06,JFMsbragag06,Stonelauga03,Philip72,Sbragagprosper06,troian05,Davies06}. 

Ou and co-workers \cite{ou04,ou05} investigated the performance of a 
microchannel with a rectangular cross section the floor of which contained a 
series of grooves aligned with the flow. This configuration resembles the
situation studied theoretically by Philip \cite{Philip72} who considered 
a variety of such arrangements. In a system of this type, the liquid surface 
becomes convex towards the gas occupying the groove due to the high pressure 
in the channel, with the radius of curvature determined by the balance of 
the pressure difference and surface tension. The importance of this effect,
which was neglected by Philip, increases with increasing pressure and may 
be expected to become significant as the channel size decreases and the 
pressure levels correspondingly increase. The purpose of the present paper 
is to calculate the correction to the effective slip length due to this 
curvature. While the correction is small for the case considered by Ou and 
co-workers, one can easily envisage practical cases in which the effect would 
 be important. Our approach is perturbative and rests on the smallness of the 
deformation of the free surface.

\section{Curvature effects on a periodic free shear pattern: formulation of the problem}

\begin{figure}
\begin{center}
\includegraphics[scale=0.5]{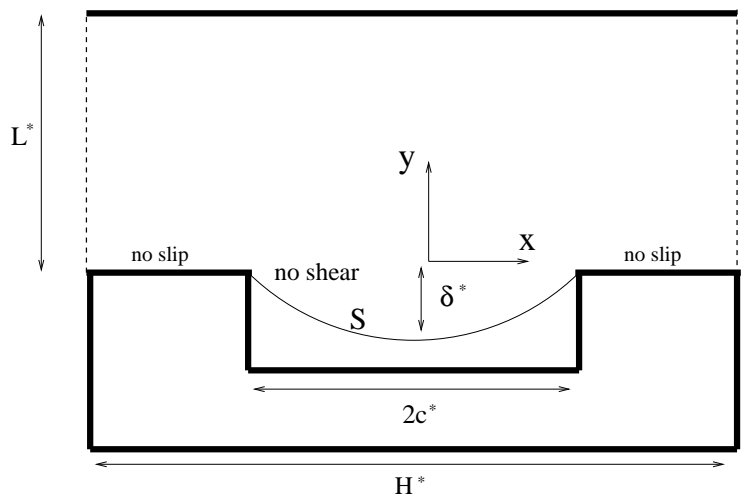}
\caption{Schematic geometry of the problem studied in this paper. The floor of 
a channel with height $L^{*}$ contains a series of regularly spaced grooves 
parallel to the flow direction. When the number of grooves is large, the 
situation can be approximated by the periodic repetition of a fundamental 
cell of width $H^*$ such as the one shown here. The width of the the groove, 
where the shear stress is essentially zero, is $2c^{*}$. For small 
deformation, the maximum penetration of the free surface $S$ into the groove is  $\delta^*=(c^{*})^{2}/2 R^{*}$ where $R^*$ is the radius of curvature of
the free surface.}
\label{fig:setup}
\end{center}
\end{figure}

We consider a liquid flowing in a rectangular channel the 
floor of which contains equally spaced slots aligned with the flow. Surface 
tension prevents the liquid from filling the slots, but the pressure in the 
channel causes the liquid free surfaces to bow into them. If there is a 
sufficient number of slots, we can consider the flow as consisting of the 
periodic repetition of equal cells similar to the one shown in figure 
\ref{fig:setup}. In this figure, the top boundary, at a distance $L^{*}$ above 
the floor, is the roof of the channel, 
the dashed vertical lines are periodicity boundaries, and the lower boundary 
of the liquid region consists of a free-slip portion along a free surface $S$, 
occupying the range $|x^*|<c^{*}$, and a no-slip portions on either side of 
it, for $c^{*} <|x^*|< H^{*}/2$, where $H^*$ is the 
dimensional cell width. Lengths will be rendered dimensionless  with respect 
to $H^*/2\pi$, so that $0\leq c <\pi$ and $L= 2\pi L^*/H^*$. It will also be
convenient to introduce a slip fraction defined as 
\be
     \xi\,=\, {2c^* \over H^*}\,=\, {c\over \pi} \, .
\label{defxi}
\ee
In fully developed conditions, only the axial velocity component $w^*$ is 
non-zero. Upon non-dimensionalizing it with respect to 
the axial pressure gradient $|dp^*/dz^*|$ and liquid viscosity $\mu$, 
$w \,=\, (2\pi/H^*)^2\mu w^*/(-dp^*/dz^*)$, this quantity satisfies
\be
\pat_{xx}w+\pat_{yy} w \,=\, -1 \, ,
\label{nseq}
\ee
where we have assumed the pressure field to be uniform over the cross section. 
The field $w$ satisfies the no-slip condition on the 
roof of the cell and periodicity conditions on its lateral boundaries:
\be
   w(x,L) \,=\, 0, \qquad \pat_x w(\pm \pi,y) \,=\, 0.
\label{rflat}
\ee
On the floor of the cell, the no-slip condition applies away from the 
groove:
\be
   w(x,0) \,=\, 0 \, \qquad {\rm for} \qquad  c < |x| <\pi 
\label{bot}
\ee
while there is no tangential stress on the free surface $S$:
\be
{\bm n}\cdot {\bm \nabla} w \,=\, 0 \qquad {\rm on} \qquad S
\label{fresh}
\ee
where ${\bm n}$ is the normal to the free surface. From a knowledge of $w$, 
one can calculate the flow rate in the channel 
\be
  Q \,=\, \int_A w(x,y) \, dA
\label{flrat}
\ee
where $A$ is the total cross-sectional area bounded by the solid walls 
and the free surface. 

One can define an effective slip length $\lambda$ 
by equating the actual flow rate $Q$ with the flow rate $Q_{eff}$ 
that would be found in a rectangular channel of height $L$ and width $2\pi$ 
with a partial-slip condition applied at the bottom wall:
\be
  w_{eff}(x,0) \,=\, \lambda \,\pat_y  w_{eff}(x,0) \, .
\label{deflam}
\ee
The flow rate in this latter case is readily calculated:
\be\label{indirect}
Q_{eff}=\frac{\pi L^3}{6}\left(1+3\frac{\lambda}{L} \right).
\ee

\section{Perturbation problem}

With the small dimensions of practical interest, the curvature of the free
surface is small and we represent it in the form
\be
y+ \epsilon \eta(x) \,=\, 0 
\label{forfrs}
\ee
in which the parameter $\epsilon$, to be estimated presently, is taken to 
be small compared to $1$. With this approximation it is easy to show that, 
correct to first order in $\epsilon$, the free-shear condition (\ref{fresh}) 
is 
\be
\tau_{yz}+\epsilon {d\eta\over dx}\, \tau_{xz}=0
\ee
where $\tau_{iz}=\partial_{i} w$, $i=x,y$, denotes the only non-vanishing 
components of the viscous stress. The smallness of $\epsilon$ suggests a 
perturbation approach to the solution of the problem and we write 
\be
 w \,=\, w^{(0)}+\epsilon w^{(1)}+ o (\epsilon) \, .
\label{tayexp}
\ee
The field $w^{(0)}$ satisfies (\ref{nseq}) while $w^{(1)}$ is harmonic. 
Both satisfy the conditions (\ref{rflat}) and (\ref{bot}) on the solid and 
periodicity boundaries while, by a Taylor-series expansion in $\epsilon$, we 
find the boundary conditions in the groove as 
\be
\pat_y w^{(0)}(x,0) \,=\, 0 \qquad {\rm for} \qquad |x| <c 
\ee
and
\be
   \pat_y w^{(1)}(x,0) \,=\, -\pat_x\left[\eta(x)\,\pat_x w^{(0)}(x,0)\right]
 \qquad {\rm for} \qquad |x| <c
\label{zershcor}
\ee
at orders $\epsilon^{0}$ and $\epsilon$ respectively.
When the expansion (\ref{tayexp}) is used to calculate the flow rate, we find 
\be\label{expansion}
  Q\,=\, Q^{(0)} + \epsilon\,\left[ Q^{(1)}_1+ Q^{(1)}_2\right] + o 
(\epsilon)
\ee
with
\be 
  Q^{(0)} \,= \int_{-\pi}^{\pi} dx \int_0^L dy w^{(0)}(x,y) 
\ee 
and 
\be
  Q^{(1)}_1 = \int_{-\pi}^{\pi} dx \int_0^L dy w^{(1)}(x,y)
\label{defq11}
\ee
\be
 Q^{(1)}_2\,=\, 
\int_{-c}^c  w^{(0)}(x,0)\,\eta(x) \, dx. 
\label{q1tot}
\ee
This second term of the ${\cal O}(\epsilon)$ contribution to the flow rate arises from the deformation of the free surface. 

In order to estimate the parameter $\epsilon$ and the shape of the deformed 
free surface it is convenient to revert temporarily to dimensional quantities. 
From the pressure in the channel $P_{ch}^*$ and in the gas occupying the slot 
under the liquid, $P_{gas}^*$, we can calculate the radius of curvature 
$R^*$ of the interface from Laplace's formula as 
\be
   R^* \,=\, {\sigma \over P_{ch}^*-P_{gas}^*}
\ee
where $\sigma$ is the surface-tension coefficient. The circle of radius $R^*$ 
passing through the points $x^*\,=\,\pm c^*, \, y^*\,=\, 0$ has the equation 
\be
  (x^*)^2+\left(y^*-\sqrt{(R^*)^2-(c^*)^2}\right)^2\,=\, (R^*)^2
\ee
from which 
\be
 y^*\,=\,\sqrt{(R^*)^2-(c^*)^2}  -\sqrt{(R^*)^2-(x^*)^2} \, \simeq \, 
-{1\over 2R^*} ((c^*)^2-(x^*)^2).
\ee
This relation can be written in the form (\ref{forfrs}) with
\be
    \epsilon^* \,=\, {1 \over 2 R^*} \, =\, {P_{ch}^*-P_{gas}^*
\over 2 \sigma}
\ee
so that 
\be
 \epsilon\,=\, {H^*\over 4\pi}\, {P_{ch}^*-P_{gas}^* \over \sigma } \, .
\ee

Upon reverting to dimensionless quantities, the free surface shape $\eta$ in (\ref{forfrs}) is thus found to be given by
\be
 \eta \,=\, c^2-x^2 \, .
\ee
The pressure in the channel falls in the direction of the flow, which will
lead to an axial variation of $R$. This effect is usually sufficiently 
slow  \cite{ou04,ou05} as not to significantly affect the assumption of 
parallel flow.

\section{Zeroth order solution: Laminar flow over a flat patterned surface}

Following the approach proposed by Philip \cite{Philip72}, we seek the solution in the form 
\be\label{wzero}
 w^{(0,L)}\,=\, -{1\over 2}y(y-L) +{1\over 2}L\tilde{w}^{(0,L)}(x,y)
\ee
where the first term is the standard two-dimensional channel flow 
profile with no-slip top and bottom walls  and  $\tilde{w}^{(0,L)}(x,y)$ is 
the correction due to the free-slip portion of the lower boundary. 
Evidently $\tilde{w}^{(0,L)}$ 
satisfies Laplace's equation with conditions  (\ref{rflat}) and (\ref{bot}) 
on the solid and periodicity boundaries, while 
\be
    \pat_y \tilde{w}^{(0,L)}(x,0) \,=-1
\qquad {\rm for} \qquad |x| <c.
\ee
Since  $\tilde{w}^{(0,L)}$ is periodic in $x$ and even, it can be expanded in 
a Fourier cosine series in $x$, after which the requirement that it be  
harmonic determines the $y$ dependence: 
\be
\tilde{w}^{(0,L)}(x,y)=\frac{a^{(0,L)}_{0}}{2}(1-\frac{y}{L})
+\sum_{n=1}^{\infty}a^{(0,L)}_{n} \,\left[ 1- e^{- 2n(L-y)}\right] 
\,e^{-ny} \, \mbox{cos} (n x)\, .
\label{wzertil}
\ee
This form ensures the absence of slip on the channel roof  $y=L$; the boundary 
conditions at the lower wall lead to the dual series equations 
\be\label{CF1}
\frac{a^{(0,L)}_{0}}{2}+\sum_{n=1}^{\infty} a^{(0,L)}_{n}\left(1-e^{-2nL}\right) \mbox{cos} (n x)= 0 \hspace{.2in} c<x<\pi
\ee
\be\label{CF2}
\frac{a^{(0,L)}_{0}}{2L}+\sum_{n=1}^{\infty} a^{(0,L)}_{n} n \left( 1+e^{-2nL}\right) \mbox{cos} (n x)=  1 \hspace{.2in} 0<x<c.
\ee
This problem was solved by Philip \cite{Philip72}, but his procedure was 
different from that adopted here and it is shown in appendix A that his 
solution is correctly recovered.  The flow rate $Q^{(0)}$ is readily calculated from (\ref{wzero}) and 
(\ref{wzertil}):
\be\label{zerothflux}
Q^{(0)} \,= \frac{\pi L^{3}}{6}\left(1+3\frac{a^{(0,L)}_{0}}{2L} \right) \,. 
\ee
For a finite height $L$, it does not seem possible to solve the dual-series 
equations  (\ref{CF1}) and (\ref{CF2}) exactly. We have calculated the 
solution numerically as explained in appendix C. The case of infinite 
depth can however be treated analytically (appendix A). In particular, 
one finds 
\be
a^{(0,\infty)}_{0}\,=\,- 4\, \mbox{log} \,\left( \mbox{cos}\,\frac{c}{2}
\right) \, 
\ee
and 
\be
{w}^{(0,\infty)}(x,0)\,=\,2\, \mbox{arccosh}\left(\frac{\mbox{cos}\,
 \frac{x}{2}}{\mbox{cos}\, \frac{c}{2}} \right) \hspace{.2in} |x|<c. 
\label{exwoinf}
\ee
A quantity that is needed to compute the first order correction is $\partial_{x} \tilde{w}^{(0,L)}(x,0)$ as it is evident from (\ref{zershcor}). This quantity can be easily expressed as
\be\label{stressseries}
\partial_{x} w^{(0,L)}(x,0)=\sum_{n=1}^{\infty} n a^{(0,L)}_{n} \mbox{sin} (nx)[e^{-2 n L}-1]
\ee
and, in  the limit $L/H \rightarrow \infty$, the series can be summed exactly to obtain  (see appendix A):
\be\label{stress1}
\partial_{x}{w}^{(0,\infty)}(x,0)=-\sqrt{2}\frac{ \mbox{sin}\,\frac{x}{2}}{\sqrt{\mbox{cos}\,x-\mbox{cos}\,c}}  \hspace{.2in} |x|<c. 
\ee

\section{First-order problem} 

We seek the solution for the first-order correction to the velocity field in 
a form similar to (\ref{wzero}): 
\be
w^{(1,L)}(x,y)=\frac{a^{(1,L)}_{0}}{2}(1-\frac{y}{L})
+\sum_{n=1}^{\infty}a^{(1,L)}_{n} \,\left[ 1- e^{- 2n(L-y)}\right] 
\,e^{-ny} \, \mbox{cos} (n x)\, .
\label{wzerti1}
\ee
According to (\ref{defq11}), the contribution to the flow rate given by the 
velocity $\frac{L}{2} w^{(1,L)}$ is 
\be
 Q^{(1)}_1\,=\, {\pi \over 4} a^{(1,L)}_{0} L^2 \, .
\label{q1cor}
\ee
The no-slip condition imposed on (\ref{wzerti1}) is expressed by
\be
\frac{a^{(0,L)}_{0}}{2}+\sum_{n=1}^{\infty} a^{(0,L)}_{n}\left(1-e^{-2nL}\right) \mbox{cos} (n x)= 0 \hspace{.2in} c<x<\pi
\label{pertns}
\ee
while the zero-shear condition (\ref{zershcor}) becomes
\be
\frac{a^{(0,L)}_{0}}{2L}+\sum_{n=1}^{\infty} a^{(0,L)}_{n} n \left( 1+e^{-2nL}\right) \mbox{cos} (n x)= \partial_{x} (\eta \partial_{x} w^{(0,L)}(x,0)) \hspace{.2in} 0<x<c
\label{perts}
\ee
with $\partial_{x} w^{(0,L)}(x,0)$ given by (\ref{stressseries}). 
Again, this dual-series system cannot be solved analytically 
except in the case of infinite depth for which we find (see appendix B)
\be
a^{(1,\infty)}_{0}=-\frac{4}{\pi} \displaystyle\int_{0}^{c}\frac{\mbox{sin}^{2} (\frac{x}{2}) (c^2-x^2) ~ dx}{\mbox{cos}(x)-\mbox{cos}(c)}=-\frac{2}{\pi} \displaystyle\int_{0}^{c}\frac{(1-\mbox{cos} (x))(c^2-x^2) ~ dx}{\mbox{cos}(x)-\mbox{cos}(c)}.
\ee
For general $L$, we have resort to a numerical solution. 

\section{Effective slip length} 
\label{effla}

\begin{center}
\begin{figure}
\includegraphics[scale=1.0]{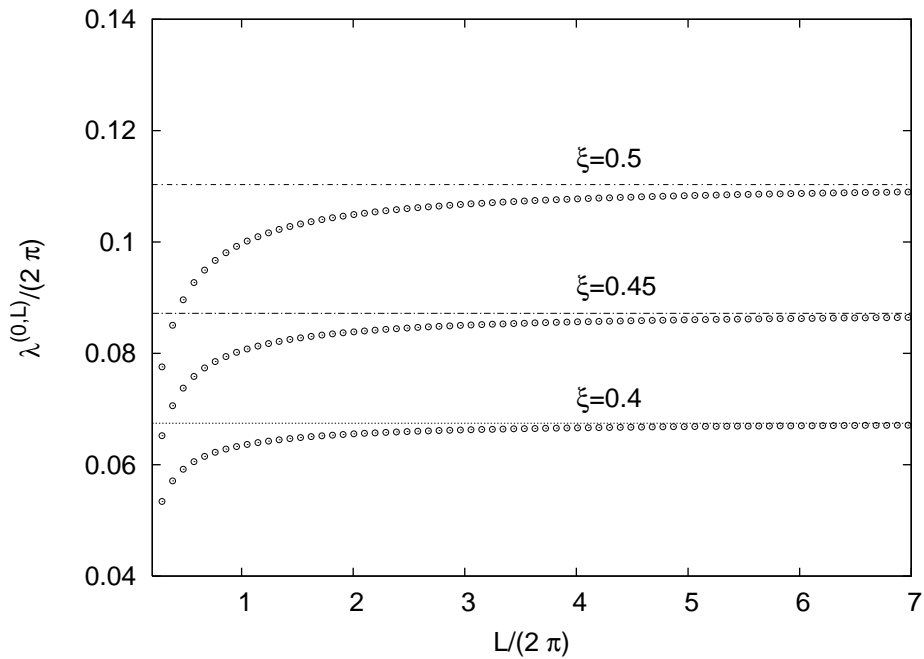}
\caption{The zero-order effective slip length $\lambda^{(0,L)}$ 
normalized by the horizontal period $2 \pi$ as a function of the channel height  $L$ also normalized by $2 \pi$ for various values of the slip fraction 
$\xi \,=\,c/\pi$; the nomenclature is defined in figure \ref{fig:setup}. 
The symbols are results obtained by solving numerically (\ref{CF1}) and 
(\ref{CF2}). The straight lines are the limit $L/(2 \pi) \rightarrow \infty$ 
given by the exact solution (\ref{EXACT}).} 
\label{fig:1}
\end{figure}
\end{center}

We expand the slip length similarly to (\ref{expansion}) as 
\be
  \lambda^{(L)}\,=\, \lambda^{(0,L)}+\epsilon \left[ \lambda^{(1,L)}_1+
 \lambda^{(1,L)}_2\right]+ o(\epsilon)
\ee
in which $\lambda^{(0,L)}$ is the contribution of the unperturbed flow 
(i.e., with a flat free-slip surface), $\lambda^{(1,L)}_1$ is the 
contribution of the perturbed velocity profile, and $\lambda^{(1,L)}_2$ 
is the contribution of the unperturbed velocity arising from the deformation 
of the flow passage. From (\ref{indirect}), (\ref{expansion}), (\ref{zerothflux}) and (\ref{q1cor}) we find 
\be
 \lambda^{(0)} \,=\frac{a_{0}^{(0,L)}}{2} 
\ee
\be
\lambda^{(1,L)}_1 \,=\, \frac{a^{(1,L)}_{0}}{2} 
\ee
\be
\lambda^{(1,L)}_2 \,= \frac{2}{\pi L^2}\int_{-c}^c  w^{(0)}(x,0)(c^2-x^2) \, dx.
\ee
Figure \ref{fig:1} is a graph of the zero-order slip length normalized 
by the width of the periodic cell: 
\be
{\lambda^{(0,L)} \over 2 \pi} \, =\, {\lambda^{(0,L^*)} \over H^*} 
\ee
as a function of the  normalized channel height 
$L/(2 \pi)\,=\,L^*/H^*$ for various values of the  slip fraction 
$\xi$. The straight lines are the 
corresponding analytical results for $L/(2 \pi) \rightarrow \infty$ given by 
(see appendix A)
\be\label{EXACT}
 \frac{\lambda^{(0,L)}}{2 \pi}\,=\, - \frac{1}{\pi}
\mbox{log} \left( \mbox{cos}\,  \frac{\pi \xi}{2}   \right).  
\ee
While convergence to this result is evident from the figure, it is also 
clear that the rate of convergence becomes slower and slower as the free-slip 
fraction increases. This feature is a consequence of the fact that, as 
follows from Eq. (\ref{deflam}) and as is explicitly shown by (\ref{EXACT}), 
the slip length diverges to infinity when $\xi\rightarrow$ 1. 

Figures \ref{fig:3} and \ref{fig:4} are a similar representation for the first-order corrections, $\lambda^{(1,L)}_1$ and $\lambda^{(1,L)}_2$, found from the numerical  solution of (\ref{pertns}) and (\ref{perts}), also for various values of  the slip fraction $\xi$. Note that the term $\lambda^{(1,L)}_2$ (see figure \ref{fig:4}) is a positive correction decreasing to zero when $L/(2 \pi) \rightarrow \infty$. On the other hand, the negative correction $\lambda^{(1,L)}_1$ is approaching a finite limit (straight lines of figure \ref{fig:3}) in the limit of large channel height where, as shown in appendix B, $\epsilon \lambda^{(1,\infty)}_{1}$ normalized to the pattern dimension can be represented as 
\be\label{F1bb}
\epsilon \frac{\lambda^{(1,\infty)}_{1}}{2 \pi} =-\frac{\delta^{*}}{H^{*}} F(\xi) \qquad \delta^*\,=\, {(c^*)^2\over 2R^*} 
\ee
with
\be\label{F2bb}
 F(\xi) = \xi\, \displaystyle\int_{0}^{1}\frac{(1-\mbox{cos} (s \pi \xi))(1-s^2) ~ ds}{\mbox{cos}(s \pi \xi)-\mbox{cos}(\pi \xi)}
\ee
A graph of this function is given in figure \ref{fig:2}. Its asymptotic 
behaviors for $\xi$ near 0 and 1 is given by 
\be
F(\xi)\,\simeq \,\frac{1}{3}\xi+\frac{\pi^2}{36}\xi^3+\frac{\pi^4}{450} 
\xi^5 +{\cal O}\left( \xi^7 \right) \qquad \xi \rightarrow 0
\ee
and 
\be
 F(\xi)\,\simeq \, \frac{2}{\pi^3 \xi^2} \frac{1}{\cos {\pi \xi \over 2}} \qquad \xi \rightarrow 1 \,. 
\ee

\section{Summary and conclusions}

Super-hydrophobic surfaces are necessary to significantly affect pressure 
gradients and facilitate liquid flow in micron-scale channels. The 
super-hydrophobic effect is realized by patterning the channel walls with 
posts or grooves so that the liquid forms a free surface and remains 
partially suspended away from the wall due to the effect of surface tension. 
While, on small scales, surface tension is a powerful force, it cannot prevent 
the free surface from bulging out becoming convex toward the gas space. This 
circumstance has two effects. In the first place, the flow passage is 
enlarged, which increases the slip length but, on the other hand, 
the velocity field is modified, which decreases it. 

\begin{center}
\begin{figure}
\includegraphics[scale=1.0]{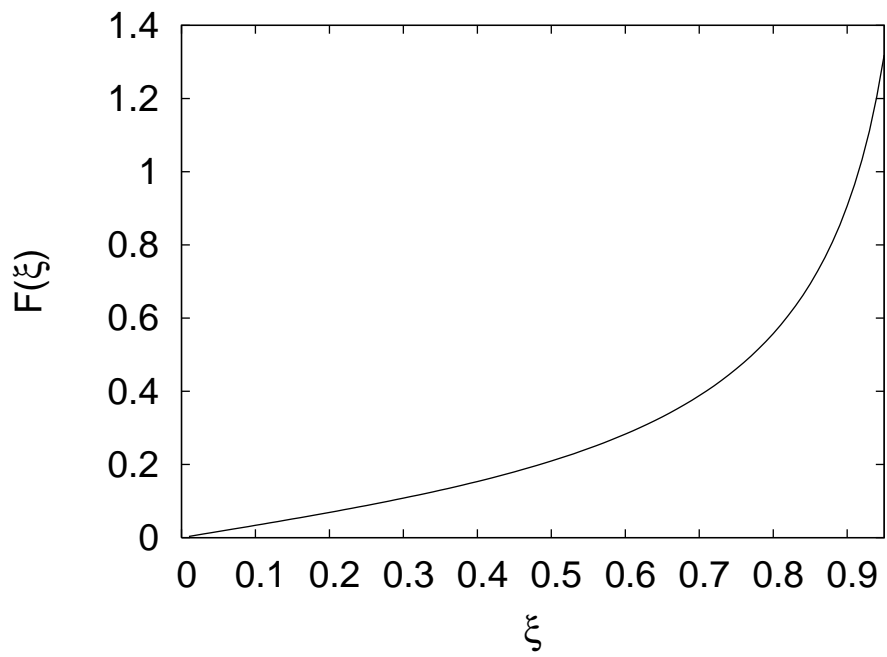}
\caption{The function $F(\xi)$ as it is defined in (\ref{F2bb}) is plotted as a function of the slip percentage $\xi$.}
\label{fig:2}
\end{figure}
\end{center}

\begin{center}
\begin{figure}
\includegraphics[scale=1.0]{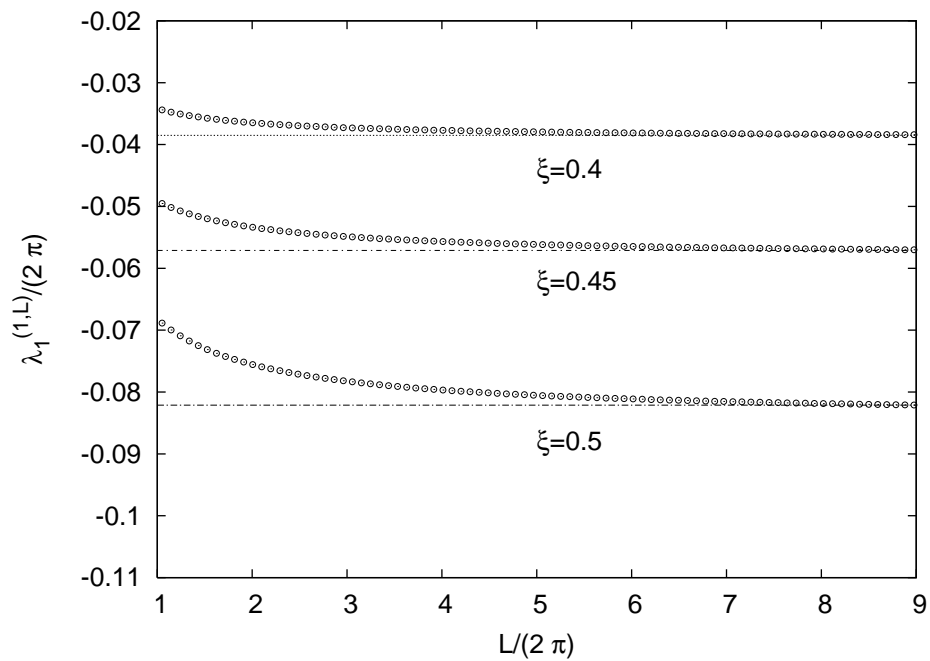}
\caption{Results for the quantity $\lambda^{(1,L)}_{1}/(2 \pi)$ as a function of the dimensionless ratio $L/(2 \pi)$ for various values of $\xi$. The symbols are results obtained by a numerical solution of (\ref{pertns}) and (\ref{perts}). The limit $L/(2 \pi) \gg 1$ corresponds to the analytical solution computed in appendix B.}
\label{fig:3}
\end{figure}
\end{center}

\begin{center}
\begin{figure}
\includegraphics[scale=1.0]{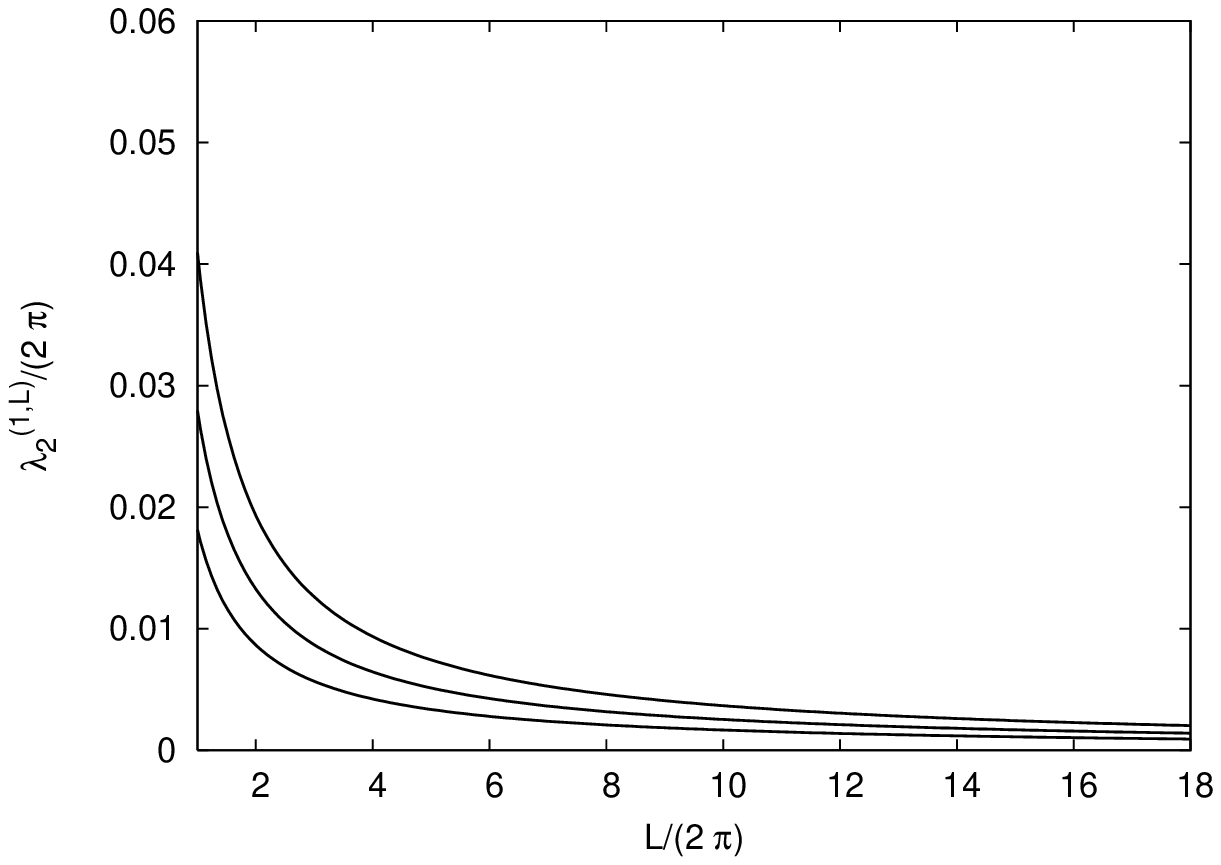}
\caption{Results for the quantity $\lambda^{(1,L)}_{2}/(2 \pi)$ as a function of the dimensionless ratio $L/(2 \pi)$ for various values of $\xi$: $\xi=0.5$ (top) $\xi=0.45$ (middle) and $\xi=0.4$ (bottom). Results are obtained by a numerical solution of (\ref{pertns}) and (\ref{perts}).}
\label{fig:4}
\end{figure}
\end{center}

In this paper we have considered one special situation of this type -- a flat 
wall with grooves parallel to the flow direction -- thus providing a natural 
generalization of the analysis carried out  by Philip \cite{Philip72} who 
assumed a flat the liquid-gas interface. We have quantified the two effects
mentioned before finding that the magnitude of the first one relative to the 
second one decreases proportionally to the ratio of the pattern width to the 
channel height. For deep channels, the more significant effect is the 
second one, given in (\ref{F1bb}) and shown in figure~\ref{fig:2}, which 
we found to give a negative contribution to the slip length. Our results 
are described in detail in section \ref{effla}. 

The magnitude of the effect that we have studied is quantified by the 
dimensionless ratio 
\be 
 \epsilon\,=\, {H^*\over 4\pi}\, {P_{ch}^*-P_{gas}^* \over \sigma } . 
\ee
The only existing data with which our theory can be compared 
are those reported by Ou and coworkers \cite{ou04}. We can estimate from 
their figure 9 that the air-water interface protrudes 2 to 4 $\mu$m below 
the channel floor. The spatial period of the grooves on their lower wall 
is $H^{*}$ = 60 $\mu$m. Our parameter $\epsilon$ is therefore of the order 
of (2-4)/60 = 0.03-0.06, which produces only a very small correction to 
the unperturbed solution. However, for periodicity patterns 
$H^*\sim $ 10 $\mu$m, with $P_{ch}^*-P_{gas}^*\sim$ 100 kPa, $\sigma \sim$ 
0.1 N/m, we have $\epsilon\sim$ 0.1, an estimate that would increase further 
with increasing area fraction $\xi$ as shown in figure \ref{fig:2}.

\section*{Acknowledgments}

M. Sbragaglia is grateful to STW (Nanoned Programme) for financial support.

\section{Appendix A}

In this appendix we solve the zeroth order problem for a channel of infinite 
depth. In this limit, the appropriate situation to consider is that of a 
linear shear flow 
over a periodic array of free-slip longitudinal strips. We write the velocity 
field as the sum of a linear shear flow  plus a perturbation 
($\tilde{w}^{(0,\infty)}$):
\be
w^{(0,\infty)}(x,y)=y+\tilde{w}^{(0,\infty)}(x,y)
\ee
with the correction expanded as
\be
\tilde{w}^{(0,\infty)}(x,y)=\frac{a^{(0,\infty)}_{0}}{2}+\sum_{n=1}^{\infty} a^{(0,\infty)}_{n} \mbox{cos} (n x) e^{-n y}.
\ee
The boundary conditions (\ref{bot}) and (\ref{fresh}) give rise to 
the dual series problem
\be
\frac{a^{(0,\infty)}_{0}}{2}+\sum_{n=1}^{\infty} a^{(0,\infty)}_{n} \mbox{cos} (n x)= 0 \hspace{.2in} c<x<\pi\ee
\be
\sum_{n=1}^{\infty} a^{(0,\infty)}_{n} n \mbox{cos} (n x)=  1 \hspace{.2in} 0<x<c.
\ee
For the problem 
\be
\frac{1}{2}a^{(0,\infty)}_{0}+\sum_{n=1}^{\infty}  a^{(0,\infty)}_{n} \mbox{cos} (nx)=0 \hspace{.2in} c<x<\pi
\ee
\be
\sum_{n=1}^{\infty} n a^{(0,\infty)}_{n} \mbox{cos} (nx)=f(x)  \hspace{.2in} 0<x<c
\ee
Sneddon (p.161) \cite{sneddon66} gives the solution 
\be\label{a0}
a^{(0,\infty)}_{0}=\frac{2}{\pi}\left[ \frac{\pi}{\sqrt{2}}\displaystyle\int_{0}^{c}h^{(0,\infty)}_{1}(t) dt \right]
\ee
\be\label{coefficients0}
a^{(0,\infty)}_{n}=\frac{2}{\pi}\left[ \frac{\pi}{2\sqrt{2}}\displaystyle\int_{0}^{c}h^{(0,\infty)}_{1}(t)[P_{n}(\mbox{cos} (t) )+P_{n-1}(\mbox{cos} (t))] dt \right] \hspace{.2in}  n=1,2,...
\ee
with $P_{n}$ Legendre polynomials. The function $h^{(0,\infty)}_{1}(t)$ is
\be\label{h}
h^{(0,\infty)}_{1}(t)=\frac{2}{\pi}\frac{d}{d t} \displaystyle\int_{0}^{t}\frac{\mbox{sin} (\frac{x}{2}) ~ dx}{\sqrt{\mbox{cos}(x)-\mbox{cos}(t)  }}
\displaystyle\int_{0}^{x}f(u)du
\ee
or, in our case, 
\be\label{h0}
h^{(0,\infty)}_{1}(t)=\frac{2}{\pi}\frac{d}{d t} \displaystyle\int_{0}^{t}\frac{x\,\mbox{sin} (\frac{x}{2}) ~ dx}{\sqrt{\mbox{cos}(x)-\mbox{cos}(t)  }} \, .
\ee

The computation of $a^{(0,\infty)}_{0}$ and of $\tilde{w}^{(0,\infty)}$ hinges 
on the knowledge of the function $h^{(0,\infty)}_{1}(t)$ which is the derivative of 
\be\label{I}
I^{(0,\infty)}(t)=\frac{2}{\pi} \displaystyle\int_{0}^{t}\frac{x ~ \mbox{sin} (\frac{x}{2}) ~ dx}{\sqrt{\mbox{cos}(x)-\mbox{cos}(t)  }}.
\ee 
This integral can be evaluated with some manipulations and the use 
of formula 3.842 of Gradshteyn \& Ryzhik \cite{GR00} with the result:
\be
I^{(0,\infty)}(t) \,=\, \frac{4 }{\sqrt{2}}\mbox{log}\left( \frac{1}
{\mbox{cos}(\frac{t}{2}) }\right)
\ee
from which 
\be
h^{(0,\infty)}_{1}(t)= \sqrt{2} \mbox{tan} \left( \frac{t}{2} \right) 
\ee
and
\be
\frac{a^{(0,\infty)}_{0}}{2}=2 \mbox{log}\left( \frac{1}{\mbox{cos}(\frac{c}{2}) }\right).
\ee
We also notice that from Sneddon (p.161) \cite{sneddon66} 
\be\label{VEL0}
\frac{1}{2}a^{(0,\infty)}_{0}+\sum_{n=1}^{\infty} a^{(0,\infty)}_{n} \mbox{cos}(n x)= \mbox{cos}{\left(\frac{x}{2}\right)}\displaystyle\int_{x}^{c}\frac{h^{(0,\infty)}_{1}(t) dt}{\sqrt{\mbox{cos}(x)-\mbox{cos}(t)   }}
\ee
that immediately leads to the velocity at $y=0$
\be
\tilde{w}^{(0,\infty)}(x,0)= \sqrt{2} \mbox{cos} \left( \frac{x}{2} \right) \displaystyle \int_{x}^{c} \frac{\mbox{tan} \left( \frac{t}{2} \right)}{\sqrt{\mbox{cos}(x)-\mbox{cos}(t)}} dt \hspace{.2in} |x|<c. 
\ee
This integral can be done exactly leading to (\ref{exwoinf}). The quantity 
$\partial_{x}w^{(0,\infty)}(x,0)$ is also readily evaluated with the result 
given in (\ref{stress1}).

\section{Appendix B}

In this appendix we solve the first order correction to the linear shear flow 
problem considered in appendix A. We write the velocity corrections as
\be
\tilde{w}^{(1,\infty)}(x,y)=\frac{a^{(1,\infty)}_{0}}{2}+\sum_{n=1}^{\infty} a^{(1,\infty)}_{n} \mbox{cos} (n x) e^{-n y} 
\ee
with the boundary conditions (\ref{bot}) for $c<x<\pi$ and 
(\ref{zershcor}) for $ 0<x<c$. Formulae (\ref{a0}) and (\ref{coefficients0})
of appendix A again apply with the function $h_1$ now given by 
\be
h^{(1,\infty)}_{1}(t)=\frac{2}{\pi} \frac{d}{d t} \displaystyle\int_{0}^{t}\frac{\mbox{sin} (\frac{x}{2}) \eta (x) \partial_{x} w^{(0,\infty)}(x,0)   ~ dx}{\sqrt{\mbox{cos}(x)-\mbox{cos}(t)  }}
\ee
from which we have 
\be\label{sliplength}
a^{(1,\infty)}_{0}\,=\,\frac{2 \sqrt{2}}{\pi} \displaystyle\int_{0}^{c}\frac{\mbox{sin} (\frac{x}{2}) \eta (x) \partial_{x} w^{(0,\infty)}(x,0)   ~ dx}{\sqrt{\mbox{cos}(x)-\mbox{cos}(c)  }}
\ee
or, upon using (\ref{stress1}) for $\partial_{x} w^{(0,\infty)}(x,0)$, 
\be
a^{(1,\infty)}_{0}\,=\,-\frac{4}{\pi} \displaystyle\int_{0}^{c}\frac{\mbox{sin}^{2} (\frac{x}{2}) (c^2-x^2) ~ dx}{\mbox{cos}(x)-\mbox{cos}(c)}=-\frac{2}{\pi} \displaystyle\int_{0}^{c}\frac{(1-\mbox{cos} (x))(c^2-x^2) ~ dx}{\mbox{cos}(x)-\mbox{cos}(c)}.
\ee
If we introduce the slip length as $\lambda^{(1,\infty)}_{1}=\frac{a^{(1,\infty)}_{0}}{2}$, when we express $\epsilon \lambda^{(1,\infty)}_{1}$ normalized to the pattern dimension we obtain:
\be\label{final}
\epsilon \frac{\lambda^{(1,\infty)}_{1}}{2 \pi}=-\frac{1}{4 \pi^2 R} \displaystyle\int_{0}^{c}\frac{(1-\mbox{cos} (x))(c^2-x^2) ~ dx}{\mbox{cos}(x)-\mbox{cos}(c)}.
\ee
If we introduce $s=x/c$, $\delta=\frac{c^2}{2 R}$ and $c=\pi \xi$ the previous expression becomes 
\be\label{F1}
\epsilon\frac{\lambda^{(1,\infty)}_{1}}{2 \pi}=-\frac{\delta}{2 \pi} F(\xi)
\ee
with
\be
\label{F2}
 F(\xi) \,=\,\xi\,  \displaystyle\int_{0}^{1}\frac{(1-\mbox{cos} (s \pi \xi))(1-s^2) ~ ds}{\mbox{cos}(s \pi \xi)-\mbox{cos}(\pi \xi)} \, .
\ee
A partial evaluation gives  
\be\label{final}
F(\xi)= \frac{2}{\pi^3 \xi^2}\,\tan\, \left( {1\over 2}\pi\xi \right)  \left(\int_{0}^{2 \pi \xi }S_{1}(\alpha) d \alpha- \pi \xi S_{1}(2 \pi \xi)  \right)  -\frac{2}{3}\xi
\ee
in which
\be
S_1(\alpha) \,=\,\sum_{n=1}^{\infty}  \frac{\mbox{sin} ( n \alpha )}{n^2}
 \,=\, {1\over 2}\, \int_0^\alpha \log {1\over 2(1-\cos (y))} \,d y.
\ee
Integration by parts leads to 
\be
 S_1(\alpha) \,=\, \alpha \left(\log {1\over 2} -
\log \,\sin \left( {1\over 2}\alpha\right) \right) + 2\,\,\int_0^{\alpha/2} x \,\, {\rm cot} \,x \,dx \, . 
\ee
The last integral cannot be evaluated in closed form.

The asymptotic results mentioned at the end of section \ref{effla} are 
readily derived from these expressions.

\section{Appendix C}

In this section we briefly sketch the procedure used to solve the dual series 
equations numerically. Both the unperturbed problem (\ref{CF1}), (\ref{CF2}) 
and the perturbed problem (\ref{pertns}), (\ref{perts}) have the general 
structure 
\be
\frac{A_{0}}{2}+\sum_{n=1}^{\infty} \Gamma_{n} A_{n} \, \mbox{cos} (n x)= 0 \hspace{.2in} c<x<\pi
\ee
\be
\sum_{n=1}^{\infty}  n \Psi_{n} A_{n} \, \,\mbox{cos} (n x)=  f(x) \hspace{.2in} 0<x<c
\ee
with $\Gamma_{n}$ and $\Psi_{n}$ generic functions of $n$. The two expressions are multiplied by $\mbox{cos} (n x)$ and integrated in  their respective domains of validity. Upon using the identity 
\be
\displaystyle\int_{0}^{c} \mbox{cos} (m x)  \mbox{cos} (n x) dx \, =\, 
{1\over 2}\pi \delta_{nm}-\displaystyle\int_{c}^{\pi} \mbox{cos} (m x) 
\mbox{cos} (n x) dx  
\ee
the result may be written as 
\be\label{eq1}
\displaystyle\int_{c}^{\pi} \frac{A_{0}}{2}  \mbox{cos} (m x) dx  +\sum_{n=1}^{\infty} \Gamma_{n} A_{n}  \left( - \displaystyle\int_{0}^{c} \mbox{cos} (m x) \mbox{cos} (n x) dx +  \frac{\pi}{2}\delta_{nm} \right) = 0 \hspace{.2in} 
\ee
\be\label{eq2}
\sum_{n=1}^{\infty} n \Psi_{n} A_{n} \displaystyle\int_{0}^{c}  \mbox{cos} (n x) \mbox{cos} (m x) d x=  \displaystyle\int_{0}^{c} \mbox{cos} (m x)  f(x) dx \hspace{.2in}. \ee
Upon adding these two relations, the result may be written in the form of
a linear system: 
\be\label{SYSTEM}
\sum_{n=0}^\infty M_{n,m} A_{n}=B_{m}
\ee
where
\be
M_{0,m}=\frac{1}{2}\displaystyle\int_{c}^{\pi} \mbox{cos} (m x) dx  
\ee
\be
M_{n,m}=(n\Psi_{n}-\Gamma_{n}) \displaystyle\int_{0}^{c} \mbox{cos} (m x) \mbox{cos} (n x) dx +  \frac{\pi}{2}\delta_{nm}\Gamma_{n}  \hspace{.2in} (n>1)
\ee
\be
B_{m}= \displaystyle\int_{0}^{c} \mbox{cos} (m x)  f(x) dx
\ee
The linear system (\ref{SYSTEM}) has been truncated and reduced to a 
$N \times N$ matrix and is then found to converge upon truncation refinement.

\newpage





\begin{thebibliography}{}






 
\bibitem{laugarev06} E. Lauga, M. Brenner and H. Stone, ``Microfluidics: The no slip boundary condition''in  {\it Handbook of experimental fluid mechanics} (Springer, 2006).

\bibitem{cheng02} J. T. Cheng and N. Giordano, ``Fluid flows through nanometer scale-channels'', {\it Phys. Rev. E} {\bf 65}, 0312061 (2002).

\bibitem{vino05} O. Vinogradova, G.E. Yabukov: ``Dynamic Effects on Force Measurements. 2. Lubrication and the Atomic Force Microscope'' {\it Langmuir} {\bf 19},  1227-1234 (2005).

\bibitem{zhu02} Y. Zhu and S. Granick,''Limits of the hydrodynamic no-slip boundary condition'', {\it Phys. Rev. Lett.} {\bf 88}, 106102 (2002).

\bibitem{barrat99} J.-L. Barrat and L. Boquet ,''Large slip effect at a non-wetting fluid-solid interface '' {\it Phys. Rev. Lett.} {\bf 82},  4671-4674  (1999).

\bibitem{Cottin04} C. Cottin-Bizonne, C. Barentine, E. Charlaix, E. Boquet and J.-L. Barrat, ``Dynamics of simple liquids at heterogeneous surfaces: Molecular dynamics simulations and hydrodynamic description'', {\it Eur. Phys. Jour. E} {\bf 15}, 427-438 (2004).

\bibitem{PRLsbragag06} R. Benzi, L. Biferale, M. Sbragaglia, S. Succi and F. Toschi,''On the roughness-hydrophobicity coupling in micro- and  nano-channel flows'', {\it Phys. Rev. Lett.} submitted (2006).

\bibitem{hotai98} C.-M. Ho and Y.-C. Tai, ``Micro-electro-mechanical systems (MEMS) and fluid flows'', {\it Annu. Rev. Fluid. Mech.} {\bf 30}, 579-612 (1998).

\bibitem{tabebook03} P. Tabeling, {\it Introduction a la microfluidique} (Belin, Paris, 2003).

\bibitem{ou04} J. Ou, B. Perot and J. Rothstein,''Laminar Drag reduction in microchannels using ultra-hydrophobic surfaces'', {\it Phys. Fluids} {\bf 16}, 4635 (2004).

\bibitem{ou05} J. Ou and J. Rothstein,''Direct velocity measurements of the flow past drag-reducing ultra-hydrophobic surfaces'' {\it Phys. Fluids} {\bf 17}, 103606 (2005).

\bibitem{tabe06} P. Joseph, C. Cottin-Bizonne, J.-M. Benoit, C. Ybert, C. Journet, P. Tabeling and  L. Bocquet,''Slippage of water past superhydrophobic carbon nanotubes forests in microchannels'' {\it Phys. Rev. Lett.} {\bf 97}, 156104 (2006).

\bibitem{bico99} J. Bico, C. Marzolin and D. Quere,''Pearl drops'' {\it Europhys. Lett.} {\bf 47}, 220-226 (1999).

\bibitem{oner00} D. Oner and T.J. McCarthy,''Ultra-hydrophobic surfaces. Effects of topography length scale on wettability'' {\it Langmuir} {\bf 16}, 7777-7782 (2000).

\bibitem{wata99} K. Watanabe, Y. Udagawa and H. Udagawa, ``Drag reduction of Newtonian fluid in a circular pipe with a highly repellent wall'' {\it Jour. Fluid. Mech.}  {\bf 381}, 225-238 (1999).

\bibitem{choi06} C.-H. Choi and C.-J. Kim,''Large Slip of Aqueous Liquid Flow over a Nanoengineered Superhydrophobic Surface'' {\it Phys. Rev. Lett.} {\bf 96}, 066001 (2006).

\bibitem{JFMsbragag06} R. Benzi, L. Biferale, M. Sbragaglia, S. Succi and F. Toschi,''Mesoscopic modelling of heterogeneous boundary conditions for microchannel flows'', {\it Jour. Fluid. Mech.} {\bf 548}, 257-280 (2006).

\bibitem{Stonelauga03} E. Lauga and H. Stone, ``Effective slip length in pressure driven Stokes flow'', {\it Jour. Fluid Mech.} {\bf 489}, 55-77 (2003).

\bibitem{Philip72} J. Philip, ''Flow satisfying mixed no-slip and  no-shear conditions'', {\it Z. Angew. Math. Phys.} {\bf 23}, 353-370 (1972).

\bibitem{Sbragagprosper06} M. Sbragaglia and A. Prosperetti,''Effective velocity boundary condition at a mixed slip surface '' {\it Jour. Fluid Mech.}, submitted (2006). See physics/0607003.

\bibitem{troian05} N. V. Priezjev, A. A. Darhuber and S. M. Troian  ``Slip behavior in liquid films on surfaces of patterned wettability: Comparison between continuum and molecular dynamics simulations'' {\it Phys. Rev. E} {\bf 71}, 041608 (2005).

\bibitem{Davies06} J. Davies, D. Maynes, B. W. Webb and B. Woolford,''Laminar Flow in a microchannel with superhydrophobic walls exhibiting transverse ribs'' {\it Physics of fluids} {\bf 18}, 087110 (2006).

 
\bibitem{GR00} I. S. Gradshteyn and  I.M. Ryzhik, {\it Table of Integrals, Series, and Products}, 6th edn. San Diego: Academic Press (2000).

\bibitem{sneddon66} I.N. Sneddon, {\it Mixed boundary value problems in potential  theory}  (Amsterdam: North-Holland, 1966).











 














 




\end{thebibliography}
\end{document}